\documentclass[twocolumn,aps,superscriptaddress,showpacs]{revtex4}

\usepackage{amssymb}
\usepackage{amsmath}
\usepackage{graphicx}
\usepackage[normalem]{ulem}
\usepackage[dvips]{color}

\begin{document}

\title{Isospin- and momentum-dependent effective interactions for the baryon octet and the properties of hybrid stars}
\author{Jun Xu}
\affiliation{Cyclotron Institute, Texas A\&M University, College Station, TX 77843-3366, USA}
\author{Lie-Wen Chen}
\affiliation{Department of Physics, Shanghai Jiao Tong University, Shanghai 200240, China} \affiliation{Center of Theoretical Nuclear Physics, National Laboratory of Heavy Ion Accelerator, Lanzhou 730000, China}
\author{Che Ming Ko}
\affiliation{Cyclotron Institute and Department of Physics and Astronomy, Texas A\&M University, College Station, TX 77843-3366, USA}
\author{Bao-An Li}
\affiliation{Department of Physics and Astronomy, Texas A\&M University-Commerce, Commerce, TX 75429-3011, USA}

\begin{abstract}

The isospin- and momentum-dependent MDI interaction, which has been
extensively used in intermediate-energy heavy-ion reactions to study
the properties of asymmetric nuclear matter, is extended to include
the nucleon-hyperon and hyperon-hyperon interactions by assuming
same density, momentum and isospin dependence as for the
nucleon-nucleon interaction. The parameters in these interactions
are determined from the empirical hyperon single-particle potentials
in symmetric nuclear matter at saturation density. The extended MDI
interaction is then used to study in the mean-field approximation
the equation of state of hypernuclear matter and also the properties
of hybrid stars by including the phase transition from the
hypernuclear matter to the quark matter at high densities. In
particular, the effects of attractive and repulsive $\Sigma$N
interactions and different values of symmetry energies on the hybrid
star properties are investigated.

\end{abstract}


\maketitle

\section{Introduction}\label{introduction}

The study of the in-medium baryon-baryon effective interactions is
one of the fundamental problems in nuclear physics.  While
significant progress has been made in understanding the
nucleon-nucleon interaction in nuclear medium, our knowledge on the
hyperon-nucleon and hyperon-hyperon interactions in nuclear medium
is very limited. The latter is important for understanding the
properties of hypernuclei, the equation of state (EOS) of dense
baryonic matter, the properties of compact stars in which hyperons
are expected to appear abundantly, and strangeness production in
heavy-ion collisions. Therefore, it is of great interest to develop
an effective model for the hyperon-nucleon and hyperon-hyperon
interactions in nuclear medium.

The in-medium single-particle potential of a nucleon generally
depends not only on the nuclear density and its momentum but also on
its isospin. The isospin- and momentum-dependent MDI interaction is
such an effective nuclear interaction based on the finite-range
Gogny interaction~\cite{Das03}. This interaction has recently been
extensively used in studying the isospin and momentum-dependent
effects in nuclear reactions and the properties of neutron stars
(for a recent review see Ref.~\cite{LCK08}). In particular, using
this interaction in the Isospin-dependent Boltzmann-Uheling-Ulenback
(IBUU) transport model with the isospin-dependent in-medium
nucleon-nucleon scattering cross sections, a relatively stringent
constraint on the density dependence of the nuclear symmetry energy
at subsaturation densities was obtained from the isospin diffusion
data in intermediate-energy heavy-ion collisions
~\cite{Tsa04,Che05a,LiBA05}. The resulting symmetry energy has been
used to constrain the neutron skin thickness of heavy
nuclei~\cite{Chen05} and the properties of neutron
stars~\cite{Steiner06,Krastev08a,Krastev08b,Wen09,Newton09}
including the transition density, which separates the liquid core
from the inner crust of a neutron star, and other properties of
neutron stars by assuming that they only consist of nucleons,
electrons, and muons~\cite{XCLM09a,XCLM09b}.

In this paper, we extend the MDI interaction to include
nucleon-hyperon and hyperon-hyperon interactions. This is achieved
by assuming that the nucleon-hyperon and hyperon-hyperon
interactions have the same density and momentum dependence as the
nucleon-nucleon interaction with the interaction parameters fitted
to known experimental data at normal nuclear matter density. With
this extended MDI interaction, we then study, as an example, the
properties of hybrid stars that not only consist of appreciable
fraction of hyperons but also a possible quark matter in their dense
core~\cite{Itoh70,Collins75,Heinz86,Burgio02a,Burgio02b,Maieron04,DiToro06,Bordbar06,Pan07,Blasch07,Carroll08,Yang08,Peng08}.
For the hadron-quark phase transition, we use the Gibbs
construction~\cite{Glen92,Glen01} with the quark phase described by
a simple MIT bag model~\cite{Chodos74,Heinz86}. In particular, we
study the effects of attractive and repulsive $\Sigma$N interactions
and different symmetry energies on the properties of hybrid stars.
We note that there have been extensive studies on hybrid stars based
on various approaches, such as the relativistic mean field (RMF)
model~\cite{Schaffer93,Schaffer96,Schaffer00,Shao09,Weng09}, the
Brueckner-Hartree-Fock
theroy~\cite{Schulze95,Schulze98,Baldo98,Baldo00,Vidana00}, and
phenomenological models~\cite{Balberg97,Banik00}.

This paper is organized as follows. In Sec.~\ref{model}, we describe
the method used in extending the MDI interaction to include the
hyperon-nucleon and hyperon-hyperon interactions. The extended MDI
interaction is then used in Sec.~\ref{hadron} to study the
thermodynamical properties of hypernuclear matter and its
equilibrium conditions. The MIT bag model is used in
Sec.~\ref{quark} to describe the properties of the quark matter and
in Sec.~\ref{mixedphase} to study the hadron-quark phase transition
in dense matter. In Sec.~\ref{results}, we show and discuss the
results on the particle fractions in dense matter and its equation
of state as well as the mass-radius relation of hybrid stars in
different scenarios. A summary is given in Sec.~\ref{summary}.

\section{The MDI interaction with hyperons}\label{model}

The MDI interaction is an effective nuclear interaction with its
density and momentum dependence deduced from the phenomenological
finite-range Gogny interaction~\cite{Das03}. In the mean-field
approximation, the potential energy density in a nuclear matter of
density $\rho$ and isospin asymmetry $\delta=(\rho_n-\rho_p)/\rho$
with $\rho_n$ and $\rho_p$ being, respectively, the neutron and
proton densities, is given by
\begin{eqnarray}
V(\rho,\delta ) &=&\frac{A_{u}(x)\rho _{n}\rho _{p}}{\rho _{0}}
+\frac{A_{l}(x)}{2\rho _{0}}(\rho _{n}^{2}+\rho
_{p}^{2})+\frac{B}{\sigma +1}\frac{\rho ^{\sigma +1}}{\rho
_{0}^{\sigma }} \notag\\
&\times& (1-x\delta ^{2}) +\frac{1}{\rho _{0}}\sum_{\tau ,\tau
^{\prime }}C_{\tau ,\tau ^{\prime }} \notag\\
&\times& \int \int d^{3}pd^{3}p^{\prime }\frac{f_{\tau }(\vec{r},\vec{p}%
)f_{\tau ^{\prime }}(\vec{r},\vec{p}^{\prime
})}{1+(\vec{p}-\vec{p}^{\prime })^{2}/\Lambda ^{2}},  \label{MDIVB}
\end{eqnarray}%
where $\tau(\tau^\prime)$ is the nucleon isospin, $f_{\tau
}(\vec{r},\vec{p})$ is the nucleon phase distribution function, and
$\rho_0=0.16$ fm$^{-3}$ is the saturation density of normal nuclear
matter. Values of the parameters $A_u(x)$, $A_l(x)$, $B$, $\sigma$,
$\Lambda$, $C_l=C_{\tau,\tau}$ and $C_u=C_{\tau,-\tau}$ can be found
in Refs.~\cite{Das03,Che05a}. The parameter $x$,
related to the coefficient of the spin-exchange operator in
the Gogny/Skyrme interactions, is used to adjust the density
dependence of the symmetry energy away from the saturation density
without changing the properties of symmetric nuclear matter. At
subsaturation densities, its value has been constrained between $0$
and $-1$ from the analysis of the isospin-diffusion and neutron skin
thickness data~\cite{Tsa04,Che05a,LiBA05,Tsa09}.

\subsection{The extended MDI interaction}

To extend the MDI interaction for the nucleon-nucleon ($NN$)
interaction to include the nucleon-hyperon ($NY$) and
hyperon-hyperon ($YY$) interactions, we assume that the latter have
the same density and momentum dependence as that between two
nucleons. The potential energy density of a hypernuclear matter due
to interactions between any two baryons then has  the following
general form:
\begin{eqnarray}
V_{bb^\prime} &=& \sum_{\tau_b,\tau^\prime_{b^\prime}}\left[
\frac{A_{bb^\prime}}{2 \rho_0} \rho_{\tau_b}
\rho_{\tau^\prime_{b^\prime}} + \frac{A^\prime_{bb^\prime}}{2
\rho_0} \tau_b \tau_{b^\prime} \rho_{\tau_b}
\rho_{\tau^\prime_{b^\prime}}\right. \notag\\
&+& \left.\frac{B_{bb^\prime}}{\sigma+1}
\frac{\rho^{\sigma-1}}{\rho^\sigma_0}(\rho_{\tau_b}
\rho_{\tau^\prime_{b^\prime}} - x \tau_b \tau_{b^\prime}
\rho_{\tau_b} \rho_{\tau^\prime_{b^\prime}})\right. \notag\\
&+& \left.\frac{C_{{\tau_b},{\tau^\prime_{b^\prime}}}}{\rho_0} \int
\int d^{3}pd^{3}p^{\prime }\frac{f_{\tau_b}(\vec{r},\vec{p}%
)f_{\tau^\prime_{b^\prime}}(\vec{r},\vec{p}^{\prime
})}{1+(\vec{p}-\vec{p}^{\prime })^{2}/\Lambda ^{2}}\right],
\label{Vbb}
\end{eqnarray}
where $b$ ($b^\prime$) denotes the baryon octet included in the
present study, i.e., $N$, $\Lambda$, $\Sigma$, and $\Xi$. We use the
conventions that $\tau_N=-1$ for neutron and $1$ for
proton~\cite{footnote}, $\tau_\Lambda=0$ for $\Lambda$,
$\tau_\Sigma=-1$ for $\Sigma^-$, $0$ for $\Sigma^0$ and $1$ for
$\Sigma^+$, and $\tau_\Xi=-1$ for $\Xi^-$ and $1$ for $\Xi^0$. In
the above, the total baryon density is now given by $\rho = \sum_b
\sum_{\tau_b} \rho_{\tau_b}$, and $f_{\tau_b}({\vec r},{\vec p})$ is
the phase-space distribution function of particle species $\tau_b$.
The interaction parameters are denoted by $A_{bb^\prime}$,
$A^\prime_{bb^\prime}$, $B_{bb^\prime}$, and
$C_{\tau_b,\tau_b^\prime}$. If there are only nucleons, we can
rewrite $A_{NN}=(A_l+A_u)/2$, $A^\prime_{NN}=(A_l-A_u)/2$,
$B_{NN}=B$, and $C_{{\tau_N},{\tau^\prime_{N}}}=C_l$ for
$\tau_N=\tau^\prime_{N}$ and $=C_u$ for $\tau_N \ne
\tau^\prime_{N}$, which then reduce to the original parameters in
the MDI interaction for nucleons~\cite{Das03,Che05a}. The parameter
$x$ is again used to model the isospin effect on the interaction
between two baryons, and its value is taken to be $0$ or $-1$ in the
present study for all baryon pairs.

The single-particle potential for a baryon of species $\tau_b$ in a
hypernuclear matter can then be obtained from the total potential
energy density of  the hypernuclear matter, given by $V_{HP}
=(1/2)\sum_{b,b^\prime} V_{bb^\prime}$, as
\begin{eqnarray}
U_{\tau_b}(p) &=& \frac{\delta}{\delta \rho_{\tau_b}}
V_{HP} \notag\\
&=&\sum_{b^\prime(b^\prime \ne
b)}\sum_{\tau^\prime_{b^\prime}}\left[ \frac{A_{bb^\prime}}{2
\rho_0} \rho_{\tau^\prime_{b^\prime}} +
\frac{A^\prime_{bb^\prime}}{2 \rho_0} \tau_b \tau^\prime_{b^\prime}
\rho_{\tau^\prime_{b^\prime}} \right.\notag\\
&+& \left.\frac{B_{bb^\prime}}{\sigma+1}
\frac{\rho^{\sigma-1}}{\rho^\sigma_0} (\rho_{\tau^\prime_{b^\prime}}
- x \tau_b \tau^\prime_{b^\prime} \rho_{\tau^\prime_{b^\prime}}) +
\frac{C_{\tau_b,\tau^\prime_{b^\prime}}}{\rho_0}
\right.\notag\\
&\times& \left. \int d^{3}p^{\prime }
\frac{f_{\tau^\prime_{b^\prime}}(\vec{r},\vec{p}^{\prime
})}{1+(\vec{p}-\vec{p}^{\prime })^{2}/\Lambda ^{2}} \right]
+\sum_{\tau^\prime_{b}} \left[ \frac{A_{bb}}{\rho_0}
\rho_{\tau^\prime_{b}} \right.\notag\\
&+& \left.\frac{A^\prime_{bb}} {\rho_0} \tau_b \tau^\prime_{b}
\rho_{\tau^\prime_{b}} + \frac{2 B_{bb}}{\sigma+1}
\frac{\rho^{\sigma-1}}{\rho^\sigma_0}( \rho_{\tau^\prime_{b}} - x
\tau_b \tau^\prime_{b} \rho_{\tau^\prime_{b}}) \right.\notag\\
&+& \left.\frac{2 C_{\tau_b,\tau^\prime_{b}}}{\rho_0} \int
d^{3}p^{\prime } \frac{f_{\tau^\prime_{b}}(\vec{r},\vec{p}^{\prime
})}{1+(\vec{p}-\vec{p}^{\prime })^{2}/\Lambda ^{2}} \right] \notag\\
&+& \sum_{b^\prime,b^\prime{^\prime} } \left[ B_{b^\prime
b^\prime{^\prime}}
\frac{\sigma-1}{\sigma+1}\frac{\rho^{\sigma-2}}{\rho^\sigma_0}
\right.\notag\\
&\times& \left.\sum_{\tau_{b^\prime}}
\sum_{\tau^\prime_{b^\prime{^\prime}}} (\rho_{\tau_{b^\prime}}
\rho_{\tau^\prime_{b^\prime{^\prime}}} - x \tau_{b^\prime}
\tau^\prime_{b^\prime{^\prime}} \rho_{\tau_{b^\prime}}
\rho_{\tau^\prime_{b^\prime{^\prime}}}) \right]. \label{Ubb}
\end{eqnarray}

For the interaction parameters $A_{bb^\prime}$,
$A^{\prime}_{bb^\prime}$, $B_{bb^\prime}$ and
$C_{{\tau_b},{\tau^\prime_{b^\prime}}}$ that involve hyperons, they
can in principle be determined from the nucleon-hyperon and
hyperon-hyperon interactions in free space. Because of the lack of
hyperon-nucleon scattering experiments, information on the
hyperon-nucleon interactions has been mainly obtained from the
hyperon single-particle potentials extracted empirically from
studying $\Lambda$~\cite{Has06} as well as
$\Sigma$~\cite{Dover89,Bart99} and $\Xi$~\cite{Dover83} production
in nuclear reactions. Although this has led to extensive studies of
the hyperon-nucleon interaction in the
past~\cite{Maessen89,Reuber94,Dabrowski99}, the isospin and momentum
dependence of the hyperon-nucleon in-medium interactions is still
not very well determined, and the situation is even worse for
hyperon-hyperon interactions. We therefore assume in the present
study that $A_{bb^\prime}$, $A^{\prime}_{bb^\prime}$,
$B_{bb^\prime}$ and $C_{{\tau_b},{\tau^\prime_{b^\prime}}}$ are all
proportional to corresponding ones in the nucleon-nucleon
interaction. Specifically, for $A_{bb^\prime}$,
$A^{\prime}_{bb^\prime}$ and $B_{bb^\prime}$, we assume
\begin{eqnarray}
A_{bb^\prime} &=& f_{bb^\prime} A_{NN}, \notag\\
A^\prime_{bb^\prime} &=& f_{bb^\prime} A^\prime_{NN}, \notag\\
B_{bb^\prime} &=& f_{bb^\prime} B_{NN},
\end{eqnarray}
and for $C_{{\tau_b},{\tau^\prime_{b^\prime}}}$ we assume
\begin{displaymath}
C_{{\tau_b},{\tau^\prime_{b^\prime}}} = \left\{\begin{array}{ll}
f_{bb^\prime} \frac{C_l+C_u}{2} &({\tau_b} ~\text{or}~ {\tau^\prime_{b^\prime}} = 0),\\
f_{bb^\prime} C_l &({\tau_b} = {\tau^\prime_{b^\prime}} \ne 0),\\
f_{bb^\prime} C_u &({\tau_b} \ne {\tau^\prime_{b^\prime}} \ne
0),\end{array} \right.
\end{displaymath}
with $\Lambda$ and $\Sigma^0$ treated differently.

\begin{table*}[tbp]
\caption{{\protect\small Parameters for the MDI-Hyp-A and MDI-Hyp-R
interactions with $x=0$ and $x=-1$. All except $\sigma$ are in units
of MeV. $A_{N\Sigma}^\prime(A)$ and $B_{N\Sigma}(A)$ are for the
MDI-Hyp-A interaction, and $A_{N\Sigma}^\prime(R)$ and
$B_{N\Sigma}(R)$ are for the MDI-Hyp-R interaction. Other parameters
are the same for both interactions.}} \label{para} \centering
\begin{tabular}{ccccccccccccc}
\hline\hline $A_{NN}$ & $A_{N\Lambda}$ & $A_{N\Sigma}$ & $A_{N\Xi}$
& $A_{\Lambda\Lambda}$ & $A_{\Lambda\Sigma}$ & $A_{\Lambda\Xi}$ &
$A_{\Sigma\Sigma}$ &
$A_{\Sigma\Xi}$ & $A_{\Xi\Xi}$ & $\Lambda$ & $\sigma$\\
$$ -108.28 & -108.28 & -108.28 & -79.04 & -68.21 & -135.34 & -135.34 & -53.05 & -108.28 & -57.39 & 263.04 & 4/3\\
\hline
$x$ & $A_{NN}^\prime$ & $A_{N\Sigma}^\prime(A)$ & $A_{N\Sigma}^\prime(R)$ & $A_{N\Xi}^\prime$ & $A_{\Sigma\Sigma}^\prime$ & $A_{\Sigma\Xi}^\prime$ & $A_{\Xi\Xi}^\prime$ \\
$$ 0 & -12.29 & -12.29 & -28.65 & -8.98 & -6.02 & -12.29 & -6.52\\
$$ -1 & -103.45 & -103.45 & -241.04 & -75.52 & -50.69 & -103.45 & -54.83 \\
\hline
 $B_{NN}$ & $B_{N\Lambda}$ & $B_{N\Sigma}(A)$ &
$B_{N\Sigma}(R)$ & $B_{N\Xi}$ &  $B_{\Lambda\Lambda}$ & $B_{\Lambda\Sigma}$ & $B_{\Lambda\Xi}$ & $B_{\Sigma\Sigma}$ & $B_{\Sigma\Xi}$ & $B_{\Xi\Xi}$\\
$$ 106.35 & 106.35 & 106.35 & 247.80 & 77.64 & 67.00 & 132.94 & 132.94 & 52.11 & 106.35 & 56.37 \\
\hline
 $C_{\tau_N,\tau_N}$ & $C_{\tau_N,-\tau_N}$ &
$C_{\tau_N,\tau_\Sigma}$ & $C_{\tau_N,-\tau_\Sigma}$ &
$C_{\tau_N,\tau_\Xi}$ & $C_{\tau_N,-\tau_\Xi}$ &
$C_{\tau_\Sigma,\tau_\Sigma}$ & $C_{\tau_\Sigma,-\tau_\Sigma}$ &
$C_{\tau_\Sigma,\tau_\Xi}$ & $C_{\tau_\Sigma,-\tau_\Xi}$ &
$C_{\tau_\Xi,\tau_\Xi}$ &
 $C_{\tau_\Xi,-\tau_\Xi}$ \\
$$ -11.70 & -103.40  & -11.70 & -103.40 & -8.54 &
-75.48 & -5.73 & -50.67 & -11.70 & -103.40 & -6.20 & -54.80\\
\hline $C_{N\Lambda}$ & $C_{N\Sigma^0}$ &  $C_{\Lambda\Lambda}$ &
$C_{\Lambda\Sigma}$ & $C_{\Lambda\Xi}$ & $C_{\Sigma^0\Sigma}$ &
 $C_{\Sigma^0\Xi}$\\
$$  -57.55 & -57.55 & -36.26 & -71.94 & -71.94 & -28.20 & -57.55\\
\hline\hline
\end{tabular}%
\end{table*}

We determine the values of $f_{bb^\prime}$ by fitting the empirical
potential $U^{(b^\prime)}_b$ of baryon $b$ at rest in a medium
consisting of baryon species $b^\prime$. For hyperons in symmetric
nuclear matter at saturation density, their potentials are
\begin{equation}\label{fitlambda}
U^{(N)}_\Lambda(\rho_N^{} = \rho_0^{}) = - 30 ~\text{MeV}
\end{equation}
for the $\Lambda$ potential from the
analysis of $(\pi^+, K^+)$ and $(K^-, \pi^-)$ reactions on
nuclei~\cite{Millener88,Chrien89} and
\begin{equation}\label{fitXi}
U^{(N)}_\Xi(\rho_N^{} = \rho_0^{}) = - 18 ~\text{MeV}
\end{equation}
for the $\Xi$ potential from the
analysis of $(\Xi,~^4_{\Lambda}H)$~\cite{Aok95} and $(K^-,
K^+)$~\cite{Fukuda98,Khaustov00} reactions. This leads to the values
$f_{N\Lambda} = 1$ and $f_{N\Xi} = 0.73$. For the $\Sigma$ hyperon,
its potential was taken to be attractive in earlier
studies~\cite{Dover89}, but more recent analysis indicate that it is
repulsive in the nuclear
medium~\cite{Batty94,Mares95,Noumi02,Har05,Friedman07}. In the
present work, we consider both the attractive and repulsive cases
\begin{eqnarray}\label{fitsigma}
U^{(N)}_\Sigma(\rho_N^{} = \rho_0^{}) &=& \pm 30 ~\text{MeV}.
\end{eqnarray}
By setting $f_{N\Sigma}=1$ we get an attractive $\Sigma$N
interaction, and this is called MDI-Hyp-A in the following. To
obtain a repulsive $\Sigma$N interaction, called MDI-Hyp-R in the
following, we adjust the values of positive and negative terms in
the single-particle potential by setting $B_{N\Sigma} = 2.33 B_{NN}$
and $A_{N\Sigma}^\prime = 2.33 A_{NN}^\prime$ without changing other
parameters as in the case of MDI-Hyp-A interaction. A similar method
of changing an attractive $\Sigma$N interaction to a repulsive one
was used in the RMF calculation~\cite{Schaffer96} by changing the
coupling constants of $\omega$ and $\rho$ mesons. For the
hyperon-hyperon interaction, the parameters are fitted according
to~\cite{Schaffer94}
\begin{equation}
U^{(Y^\prime)}_Y(\rho_{Y^\prime}^{} = \rho_0^{}) \sim
-40~\text{MeV},
\end{equation}
which gives the strength of the hyperon-hyperon interactions as
$f_{\Lambda\Lambda}=0.63$, $f_{\Lambda\Sigma}=1.25$,
$f_{\Lambda\Xi}=1.25$, $f_{\Sigma\Sigma}=0.49$, $f_{\Sigma\Xi}=1$,
and $f_{\Xi\Xi}=0.53$. Detailed values of the parameters are listed
in Table.~\ref{para}. These parameterizations can be viewed as a
baseline for studying the properties of the hypernuclear matter, and
more sophisticated treatments are left for
future work after the in-medium properties of hyperons are better
understood. It will be shown in the following that many interesting
results can already be obtained even with these simple
parameterizations.

\subsection{Single-particle potentials}

An important quantity related to the interaction of a particle in
nuclear medium is its single-particle potential as given by
Eq.~(\ref{Ubb}), which is also needed later in our study of the
properties of neutron stars.  The single-particle potential of a
particle depends not only on the density of the medium but also on
the momentum of the particle. In this section, we show and discuss
the single-particle potentials of both nucleons and hyperons in
nuclear matter obtained from the extended MDI interaction.

\begin{figure}[ht]
\centerline{\includegraphics[scale=0.9]{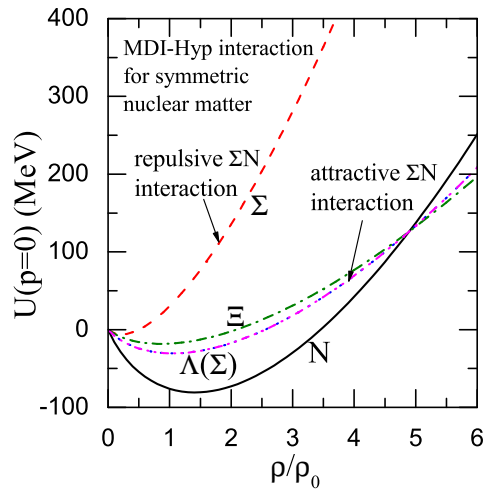}} \caption{(Color
online) Single-particle potentials for particles at rest in
symmetric nuclear matter as functions of density.} \label{SPPrhoH}
\end{figure}

We first show in Fig.~\ref{SPPrhoH} the single-particle potential of
a particle at rest in symmetric nuclear matter as a function of
density. Although the nucleon potential is more attractive at normal
density than those of hyperons, it becomes more repulsive than the
hyperon potentials above about 5 times normal nuclear density,
including the $\Sigma$ potential that is attractive at normal
density. For the $\Sigma$ potential that is repulsive at normal
density, it becomes more repulsive as the density increases and
becomes slightly attractive only at very low densities. Comparing
our results with those from other models given in Ref.~\cite{Dap09}
(and references therein), the single-particle potentials of
$\Lambda$ and $\Sigma$ are close to those from the chiral effective
field theory~\cite{Pol06}, but more repulsive than those based on
the G-matrix calculations using the soft core Nijmegen model or the
J$\rm{\ddot{u}}$lich meson-exchange model for the free
hyperon-nucleon interactions, particularly at high densities.

\begin{figure}[ht]
\centerline{\includegraphics[scale=0.9]{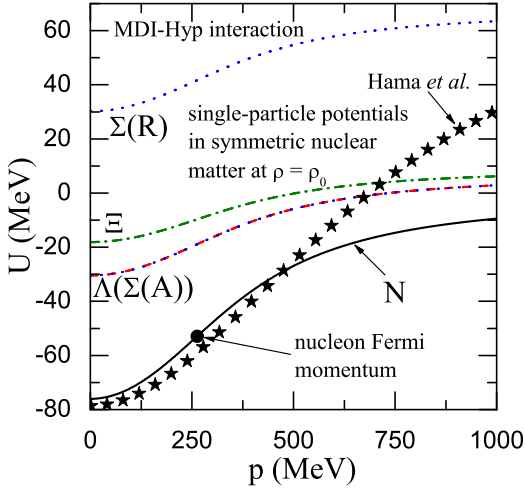}} \caption{(Color
online) Single-particle potentials in symmetric nuclear matter at
saturation density $\rho_0$ as functions of particle momentum. The
Schr$\ddot{o}$dinger equivalent potential obtained by Hama {\it et
al.}~\cite{Hama90,Coo93} from the nucleon-nucleus scattering data is
shown by stars for comparison. $\Sigma$(A) and $\Sigma$(R) are for
the MDI-Hyp-A and MDI-Hyp-R interactions, respectively. }
\label{SPPrho0H}
\end{figure}

The single-particle potential of a particle from the extended MDI
interaction also depends on its momentum. Figure~\ref{SPPrho0H}
shows the momentum dependence of the single-particle potentials for
both nucleons and hyperons in symmetric nuclear matter at saturation
density. Again, results using both attractive and repulsive
$\Sigma$N interactions (i.e., MDI-Hyp-A and MDI-Hyp-R) are shown for
comparison. Also indicated in the figure is the Fermi momentum of
nucleons. For nucleons, the single-particle potential from the MDI
interaction is consistent with the Schr$\ddot{o}$dinger equivalent
potential obtained by Hama {\it et al.} from the nucleon-nucleus
scattering data~\cite{Hama90,Coo93} up to the nucleon momentum of
$500$ MeV. For hyperons, the momentum dependence of their
single-particle potentials from the extended MDI interaction is
similar to that obtained from the G-matrix calculations based on the
free Nijmegen $NY$ interaction~\cite{Baldo98}. Both show an increase
with increasing momentum and are similar at low momenta as both are
constrained by available experimental data. They are, however,
slightly different at high momenta. The momentum dependence of $NY$
and $YY$ interactions thus remains an open question, especially at
high momenta. Also, the density dependence of the high-momentum
behavior of the mean-field potential is poorly known. Since the
maximum Fermi momenta of nucleons and hyperons reached in hybrid
stars are not very high, the incorrect nucleon potential and the
uncertainty of hyperon potentials at high momenta given by the
extended MDI interaction are thus not expected to affect
significantly the properties of cold hypernuclear matter considered
in the present study.

\begin{figure}[ht]
\centerline{\includegraphics[scale=0.9]{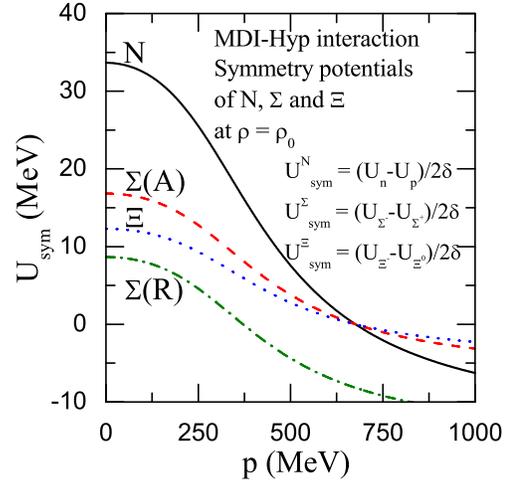}} \caption{(Color
online) Symmetry potentials of nucleon, $\Sigma$ and $\Xi$ in
asymmetric nuclear matter at saturation density $\rho=\rho_0$.
$\Sigma(A)$ and $\Sigma(R)$ are for the MDI-Hyp-A and MDI-Hyp-R
interactions, respectively.} \label{Usymrho0H}
\end{figure}

It is known that both proton and neutron single-particle potentials
in asymmetric nuclear matter of unequal proton and neutron densities
are approximately linear in the isospin asymmetry $\delta$ of the
matter. Whether this is also the case for hyperons is not clear in
the literature. For the extended MDI interaction introduced in the
present study, such a linear dependence in isospin asymmetry,
however, also holds for $\Sigma$ and $\Xi$ hyperons. The
single-particle potential of a particle in asymmetric nuclear matter
can thus be written in general as $U_{\tau_b}(\rho,\delta)\approx
U_{\tau_b}(\rho,\delta=0)-\tau_b U^{b}_{\rm sym}(\rho) \delta$ in
terms of the symmetry potential $U^b_{\rm sym}(\rho)$, defined by
$U^N_{\rm sym}(\rho)=(U_n(\rho,\delta)-U_p(\rho,\delta))/2\delta$,
$U^{\Sigma}_{\rm
sym}(\rho)=(U_{\Sigma^-}(\rho,\delta)-U_{\Sigma^+}(\rho,\delta))/2\delta$
and $U^{\Xi}_{\rm
sym}(\rho)=(U_{\Xi^-}(\rho,\delta)-U_{\Xi^0}(\rho,\delta))/2\delta$
for the nucleon, $\Sigma$ and $\Xi$ hyperons, respectively. From the
single-particle potentials of nucleons, $\Sigma$ and $\Xi$ hyperons
in asymmetric nuclear matter at normal nuclear matter density and of
isospin asymmetry $\delta=0.2$, we have calculated their symmetry
potentials using above definitions. In Fig.~\ref{Usymrho0H}, the
momentum dependence of these symmetry potentials are compared. All
symmetry potentials are seen to decrease with increasing momentum.
At zero momentum, the symmetry potentials of nucleon and $\Xi$ are
about $34$ MeV and $12$ MeV, respectively, and for the $\Sigma$
hyperon they are $17$ MeV for an attractive $\Sigma$N interaction
and $9$ MeV for a repulsive one.

\begin{figure}[h]
\centerline{\includegraphics[scale=0.9]{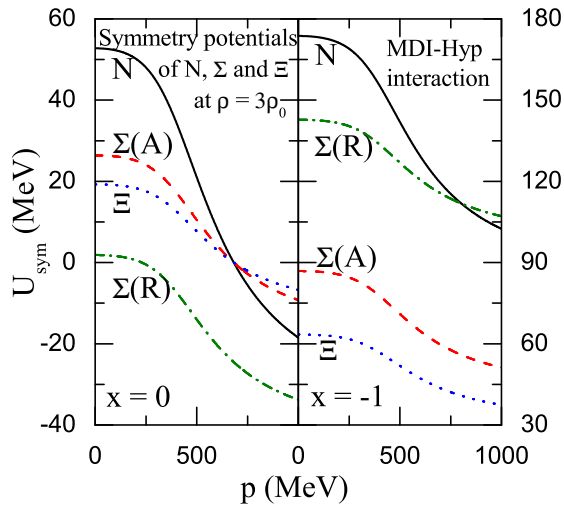}} \caption{(Color
online) Symmetry potentials of nucleon, $\Sigma$ and $\Xi$ in
asymmetric nuclear matter at density $\rho=3\rho_0$. $\Sigma(A)$ and
$\Sigma(R)$ are for the MDI-Hyp-A and MDI-Hyp-R interactions,
respectively. Note that different scales are used for $x=0$ (left
panel) and $x=-1$ (right panel).} \label{Usym3rho0H}
\end{figure}

Although the symmetry potentials at normal nuclear density are
independent of the value of $x$, which is used in the MDI
interaction to model the stiffness of nuclear symmetry energy at
densities different from the normal density, this is not the case at
other densities. This is demonstrated in Fig.~\ref{Usym3rho0H} for
the symmetry potentials of the nucleon, the $\Sigma$ hyperon and the
$\Xi$ hyperon in asymmetric nuclear matter at density $\rho=3\rho_0$
for the two symmetry energy parameters $x=0$ and $x=-1$. For $x=0$,
the symmetry potentials at zero momentum are $53$ MeV and $19$ MeV
for the nucleon and the $\Xi$ hyperon, respectively, and for the
$\Sigma$ hyperon they are $26$ MeV with the MDI-Hyp-A interaction
and $2$ MeV with the MDI-Hyp-R interaction. These values are changed
to $174$ MeV and $63$ MeV for the nucleon and the $\Xi$ hyperon,
respectively, and for the $\Sigma$ hyperon to $87$ MeV for the
MDI-Hyp-A interaction and $143$ MeV for the MDI-Hyp-R interaction
for the case of $x=-1$. As noted in Ref.~\cite{QFLi05}, the charged
$\Sigma$ baryon ratio in heavy-ion collisions can be used as a probe
to constrain the nuclear symmetry energy (potential) at densities
larger than $3\rho_0$. It will be very interesting to see how the
symmetry potentials of $\Sigma$ and $\Xi$ hyperons in nuclear matter
affect the charged $\Sigma$ hyperon ratio and the charged $\Xi$
hyperon ratio in heavy-ion collisions induced by neutron-rich
nuclei. This is important for determining the high density behavior
of the nuclear symmetry energy using these ratios in heavy-ion
collisions. The extended MDI interaction with hyperons is therefore
useful for studying the nuclear symmetry energy (potential) at
$\rho>3\rho_0$ in the transport model calculation for heavy-ion
collisions.

\section{equilibrium conditions and thermodynamical relations in dense matter} \label{beta}

The extended MDI interaction is also useful for studying the
properties of a hybrid star that is expected to have a quark core at
high densities, a mixed phase of quarks and hadrons at moderate
densities, and a hadron phase at low densities. We review in this
section the $\beta$-equilibrium, charge neutrality, and baryon
number conservation conditions of such a matter.

\subsection{The hadron phase}\label{hadron}

In the hadron phase of a hybrid star, the matter consists of
nucleons, hyperons and leptons. For leptons, we include both
electrons and muons with their masses taken to be $0$ and $106$ MeV,
respectively.  At equilibrium, these particles satisfy following
baryon number conservation, charge neutrality and
$\beta$-equilibrium conditions:
\begin{eqnarray}
\sum_i \rho_i b_i &=& \rho,\label{bnc}\\
\sum_i \rho_i q_i &=& 0,\label{cn}\\
\mu_i &=&\mu_b^Hb_i-\mu_c^Hq_i. \label{be}
\end{eqnarray}
In the above,  $q_i$ and $b_i$ are, respectively, the charge and
baryon numbers of particle species $i$, where $i$ can be nucleons,
hyperons or leptons, and their density and chemical potential are
denoted, respectively, by $\rho_i$ and $\mu_i$. The total baryon
density of the hadron phase is denoted by $\rho$, and $\mu_b^H$ and
$\mu_c^H$ are, respectively, the baryon and charge chemical
potentials of the hadron phase.

Taking into account their interactions in the mean-field
approximation, the chemical potential of baryon species $\tau_b$ is
given by
\begin{equation}
\mu_{\tau_b} (p^{}_{F{\tau_b}}) =m_{\tau_b}+
\frac{p_{F{\tau_b}}^2}{2m_{\tau_b}} + U_{\tau_b}(p^{}_{F{\tau_b}}),
\end{equation}
where $p^{}_{F{\tau_b}}$ is their Fermi momentum, and $U_{\tau_b}$
and $m_{\tau_b}$ are, respectively, the single-particle potential
and mass of the baryon. For leptons ($l=\mu$, $e$) their chemical
potential is given by $\mu_l=(m_l^2+p_{Fl}^2)^{1/2}$ with $p^{}_{Fl}
= (3 \pi^2 \rho_l)^{1/3}$ being their Fermi momentum. The relative
abundances of various hadrons and leptons for a given total baryon
density are then obtained by solving above equations.

In terms of the densities of various particles, the total energy
density of the hadron phase can be written as
\begin{equation}
\epsilon^H = V_H + V_L,
\end{equation}
where $V_H$ and $V_L$ are the contributions from baryons and
leptons, respectively. The former can be written as
\begin{equation}
V_H = V_{HP} + V_{HK} + V_{HM},
\end{equation}
where $V_{HP}=(1/2)\sum_{b,b^\prime} V_{bb^\prime}$ is the potential
energy density of baryons with $V_{bb^\prime}$ calculated from
Eq.~(\ref{Vbb}), and $V_{HK}$ and $V_{HM}$ are, respectively, the
kinetic energy and mass contributions given by
\begin{eqnarray}
V_{HK} &=& \sum_b \sum_{\tau_b} \frac{p_{F\tau_b}^5}{10\pi^2m_{\tau_b}},\\
V_{HM} &=& \sum_b \sum_{\tau_b} \rho_{\tau_b} m_{\tau_b}.
\end{eqnarray}
The contribution $V_L$ from leptons is calculated by treating them
as a free Fermi gas, i.e.,
\begin{eqnarray}
V_L &=& V_e + V_\mu, \notag\\
V_e &=& \frac{p^4_{Fe}}{4 \pi^2}, \notag\\
V_\mu &=& \frac{1}{4 \pi^2 } \left[p^{}_{F\mu} \mu^3_\mu -
\frac{1}{2} m^2_\mu p^{}_{F\mu} \mu_\mu\right.\notag\\
&-& \left.\frac{1}{2} m^4_\mu \ln\left(\frac{p^{}_{F\mu}+\mu_\mu}{m_\mu}\right) \right]. \notag\\
\end{eqnarray}

The pressure of the hadron phase is obtained from the
thermodynamical relation
\begin{equation}
P^H = P_H + P_L = (\sum_b \sum_{\tau_b} \mu_{\tau_b} \rho_{\tau_b} -
V_H) + (\sum_l \mu_l \rho_l - V_L),
\end{equation}
where $b$ and $l$ run over all species of baryons and leptons,
respectively. We note that the thermodynamical consistency condition
\begin{equation}
P^H = \rho^2 \frac{d(\epsilon^H/\rho)}{d \rho}
\end{equation}
is satisfied.

\subsection{The quark phase}\label{quark}

As the nuclear matter density increases, such as in the core of
neutron stars, not only hyperons appear but also a quark matter
could exist~\cite{Collins75}. To take into consideration possible
transition between the hadronic matter and the quark matter, we
follow many previous studies to use in the present study the MIT bag
model~\cite{Chodos74,Heinz86} to describe the cold quark matter that
might exist in the dense core of neutron stars.

For the quark phase consisting of quarks and leptons, the baryon
number conservation and charge neutrality conditions are given by
expressions similar to Eqs.~(\ref{bnc}) and (\ref{cn}) with $i$
denoting now quarks and leptons. For the $\beta$-equilibrium
condition in the quark phase, it is given by
$\mu_i=\mu_b^Qb_i-\mu_c^Qq_i$ with $\mu_b^Q$ and $\mu_c^Q$ being the
baryon and charge chemical potentials of the quark phase,
respectively.

The total energy density and pressure of the quark phase can be calculated from
\begin{eqnarray}
\epsilon^Q = V_Q + V_L, \\
P^Q = P_Q + P_L,
\end{eqnarray}
where $V_Q$ and $P_Q$ are the energy density and pressure of quarks,
which can be calculated from the MIT bag model as described in the
following, and $V_L$ and $P_L$ are the energy density and pressure
of leptons given by same expressions as those in the hadron phase.

At zero temperature, the density $\rho_f$, chemical potential
$\mu^{}_f$, and energy density $V_f$ of quarks of flavor $f=u,d,s$
in the quark matter are given,
respectively, by
\begin{eqnarray}\label{mvq}
\rho^{}_f &=& \frac{p^3_{Ff}}{\pi^2}, \notag\\
\mu^{}_f &=& \sqrt{m^2_f+p^2_{Ff}}, \notag\\
V_ f &=& \frac{3}{4 \pi^2} \left[p^{}_{Ff} \mu_f^3 - \frac{1}{2}m^2_f p^{}_{Ff} \mu^{}_f \right.\notag\\
&-& \left.\frac{1}{2} m^4_f
\ln\left(\frac{p^{}_{Ff}+\mu^{}_f}{m^{}_f}\right) \right],
\end{eqnarray}
where $p^{}_{Ff}$ is the Fermi momentum of quarks of flavor $f$. For
the quark masses, they are taken to be $m_u=m_d=0$ and $m_s=150$
MeV. In the bag model, the energy density is modified by a bag
constant $B$, resulting in an energy density given by
\begin{equation}\label{vq}
V_Q = \sum_f V_f + B.
\end{equation}
This leads to the following pressure for the quark matter:
\begin{equation}\label{pq}
P_Q = \sum_f \mu_f \rho_f - V_Q = \rho^2 \frac{d(V_Q/\rho)}{d \rho},
\end{equation}
where $\rho$ is the total baryon density of the quark phase
\begin{equation}
\rho = \frac{1}{3} \sum_f \rho_f.
\end{equation}

\subsection{The hadron-quark phase transition}\label{mixedphase}

\begin{figure}[h]
\centerline{\includegraphics[scale=0.9]{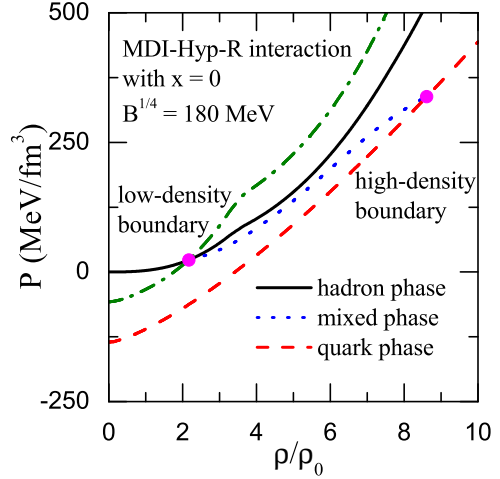}}
\caption{(Color online) Pressures as functions of baryon density $\rho$ for the
hadron phase (solid line), mixed phase (dotted line) and quark phase
(dashed line). The pressure of the quark matter that is in chemical
equilibrium with the hadronic matter as a function of the baryon
density of the hadronic matter is denoted by the dash-dotted line,
whose intersection with the solid line gives the low-density
boundary of the hadron-quark phase transition. } \label{PrhoTEST}
\end{figure}

The hadron-quark phase transition leads to a mixed phase of hadronic
and quark matters,  which is usually described by the Gibbs conditions
~\cite{Glen92,Glen01}
\begin{eqnarray}\label{Gibbs}
T^H&=&T^Q,\qquad P^H = P^Q,\notag\\
\mu_b=\mu_b^H&=&\mu_b^Q,\qquad\mu_c=\mu_c^H=\mu_c^Q.
\end{eqnarray}
The Gibbs conditions for the chemical potentials can also be expressed as
\begin{eqnarray}\label{Gibbs_mu}
&&\mu_u = \frac{1}{3}\mu_n - \frac{2}{3} \mu_e, \notag\\
&&\mu_d = \mu_s = \frac{1}{3}\mu_n + \frac{1}{3} \mu_e.
\end{eqnarray}
Since only the case of zero temperature is considered in this paper,
the first condition in Eq.~(\ref{Gibbs}) is always satisfied. To
solve above equations, we follow the method of Ref.~\cite{Glen01}.
In this method, one first calculates the pressure of the hadron
phase at a series of baryon densities. From the known chemical
potentials $\mu_n$ and $\mu_e$ of the hadronic matter at these
densities, one then calculates the pressure of the quark matter that
is in chemical equilibrium with corresponding hadronic
matter, i.e., with the quark chemical potentials $\mu_u$, $\mu_d$
and $\mu_s$ determined by the two relations in Eq.~(\ref{Gibbs_mu}).
The hadronic baryon density at which the two pressures are the same
is then the low-density boundary of the hadron-quark phase
transition. As an illustration, we show in Fig.~\ref{PrhoTEST} the
results for the case of the symmetry energy parameter $x=0$ and the
bag constant $B^{1/4}=180~{\rm MeV}$. The solid line is the density
dependence of the pressure of the hadron phase calculated using the
model described in Sec.~\ref{hadron}, while the dash-dotted line is
the pressure of the quark matter that is in chemical equilibrium
with the hadronic matter as a function of the baryon density of the
hadronic matter. The low-density boundary of the hadron-quark phase
transition is indicated by the intersection of the solid line and
the dash-dotted line.

Above the low-density boundary determined above, the dense matter
enters the mixed phase, in which the hadron phase and the quark
phase satisfy following chemical equilibrium, mechanical
equilibrium, baryon number conservation, and charge neutrality
conditions:
\begin{eqnarray}\label{mixed}
&&\mu_b b_i - \mu_c q_i = \mu_i,\label{behq}\notag\\
&&P^H = P^Q, \label{mehq}\notag\\
&&(1-Y)\sum_b\sum_{\tau_b}\rho_{\tau_b} + \frac{Y}{3}\sum_f\rho_f = \rho,\label{bnchq} \notag\\
&&(1-Y)\sum_b\sum_{\tau_b} \rho_{\tau_b} q_{\tau_b} +
\frac{Y}{3}\sum_f\rho_f q_f + \sum_l \rho_l q_l =
0,\notag\\\label{cnhq}
\end{eqnarray}
where $i$ runs over baryons, leptons and quarks, and $Y$ is the
baryon number fraction of the quark phase. We note that in the mixed
phase the total charge is zero and the leptons play an important
role in maintaining the charge neutrality and $\beta$-equilibrium
conditions. The total energy density and pressure of the mixed phase
are calculated according to
\begin{eqnarray}
\epsilon^M &=& (1-Y)V_H + YV_Q + V_L, \notag\\
P^M &=& (1-Y)P_H + YP_Q + P_L. \notag\\
\end{eqnarray}
It is obvious that we have a pure hadron phase for $Y=0$ and a pure
quark phase for $Y=1$. For the case of the symmetry energy parameter
$x=0$ and the bag constant $B^{1/4}=180~{\rm MeV}$ considered above,
the pressure of the mixed phase as a function of the baryon density
obtained from solving Eq.~(\ref{mixed}) is shown in
Fig.~\ref{PrhoTEST} by the dotted line. The mixed phase starts at
the low-density boundary of the hadron-quark phase transition and
ends at the high-density boundary of the hadron-quark phase
transition when the matter is a pure quark matter, whose pressure is
calculated using the model described in Sec.~\ref{quark} and is
shown by the dashed line in Fig.~\ref{PrhoTEST}  as a function of
the baryon density.

There is a recent study based on the Relativistic Mean-Field model
for the hadron phase and the MIT bag model for the quark phase to
compare the behavior of the mixed phase in the Gibbs construction
with that in the Maxwell construction, which does not require same
charge chemical potential for the hadronic matter and quark matter
in the mixed phase~\cite{Bhatt09}. The pressure of the mixed phase
in the Maxwell construction is found to be constant with respect to
its baryon density, which is in contrast with the increasing
pressure of the mixed phase with increasing baryon density in the
Gibbs construction as seen in Fig.~\ref{PrhoTEST}. It is worthwhile
to point out that although the charge neutrality condition in
Eq.~(\ref{cnhq}) is satisfied globally, it is violated in each
phase. A more realistic equation of state of the mixed phase can be
obtained from the Wigner-Seitz cell calculation by taking into
account the Coulomb and surface effects. The equation of state of
the mixed phase obtained from this approach lies between those from
the Gibbs and Maxwell constructions. The latter can thus be viewed
as two extreme cases corresponding to certain values of the surface
tension parameter in the Wigner-Seitz cell
calculation~\cite{Hei93,Mar08}.

\section{Results and discussions}
\label{results}

In this section, we use the extended MDI (MDI-Hyp) interaction to
study the equation of state and the relative particle fractions in
the charge neutral and $\beta$-stable hypernuclear
matter. Including the hadron-quark phase transition in the
hypernuclear matter, we further study the mass-radius relation of
hybrid stars. Results from different values of the symmetry energy
parameter $x$, the bag constant, and attractive and repulsive
$\Sigma$N interactions are compared.

\subsection{The hypernuclear matter}

\begin{figure}[h]
\centerline{\includegraphics[scale=0.9]{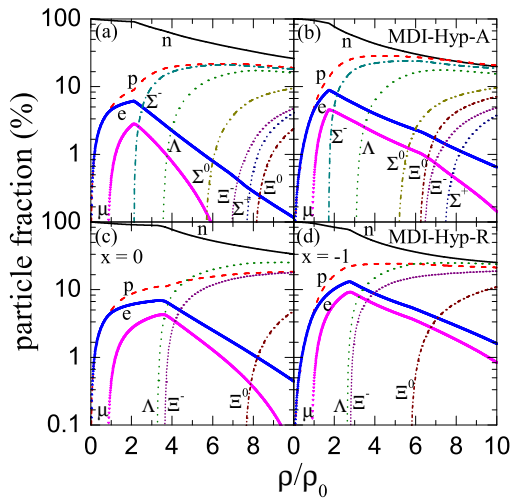}} \caption{(Color
online) Particle fractions in a hypernuclear matter for the MDI-Hyp
interaction with $x=0$ ((a) and (c)) and $x=-1$ ((b) and (d)) as
well as with attractive ((a) and (b)) and repulsive ((c) and (d))
$\Sigma$N interactions. }\label{MDIHPF}
\end{figure}

\begin{figure}[h]
\centerline{\includegraphics[scale=0.9]{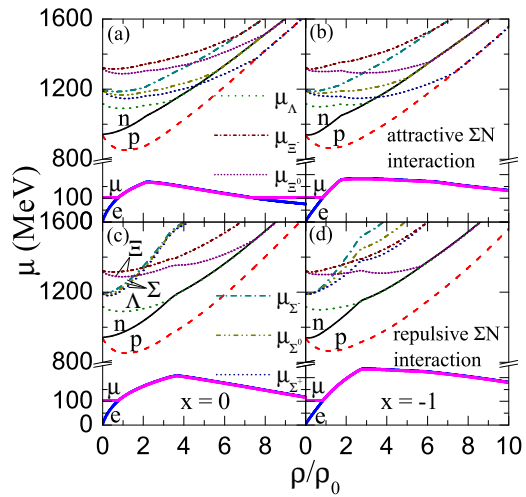}} \caption{(Color
online) Same as Fig.~\ref{MDIHPF} for the chemical potentials in a
hypernuclear matter. }\label{XMUH}
\end{figure}

Using the extended MDI interaction with hyperons (MDI-Hyp), we have
studied the particle fractions in a hypernuclear
matter as functions of the baryon density. The results are shown in
Fig.~\ref{MDIHPF} for MDI-Hyp-A ((a) with $x=0$ and (b) with $x=-1$)
and MDI-Hyp-R ((c) with $x=0$ and (d) with $x=-1$) interactions. The
density dependence of corresponding particle chemical potentials is
shown in Fig.~\ref{XMUH}. It is seen that the effect of hyperons is
more pronounced at higher densities, and the appearance of hyperons
prevents the neutron chemical potential from increasing too fast.
Similar to the results from other works, the $\Lambda$ hyperon
appears at baryon density of about $3\rho_0$. For the $\Sigma$
hyperon, the density at which it appears depends on the sign of the
$\Sigma$N interaction. For the attractive MDI-Hyp-A interaction, it
appears at about $2\rho_0$, while for the repulsive MDI-Hyp-R
interaction, it does not appear until very high densities due to the
rapid increase of the chemical potentials of $\Sigma$ hyperons with
increasing total baryon density as shown in Fig.~\ref{XMUH}, making
it difficult to satisfies the $\beta$-equilibrium condition at small
baryon densities. This result is similar to that reported in the
literature~\cite{Schaffer96,Balberg97}. Also, with the value $x=-1$,
corresponding to a stiffer symmetry energy, the fractions of leptons
are lower at subsaturation densities but higher at higher densities,
which results in a larger charge chemical potential at higher
densities and the appearance of negatively (positively) charged
hyperons at a lower (higher) density. A stiffer symmetry energy also
increases the neutron chemical potential and thus leads to a larger
total hyperon fractions at higher densities. Since the total charge
of the hypernuclear matter is conserved, an increase in the fraction
of negatively charged hyperon results in an increasing fraction of
protons and decreasing fraction of leptons. We note that for the
MDI-Hyp-R interaction and at the density $10\rho_0$, about $50\%$ of
the particle fraction is made up by hyperons, and the fraction of
$\Lambda$ particles is larger than that of protons at high
densities.

\begin{figure}[h]
\centerline{\includegraphics[width=3.2in,height=3.2in,angle=0]{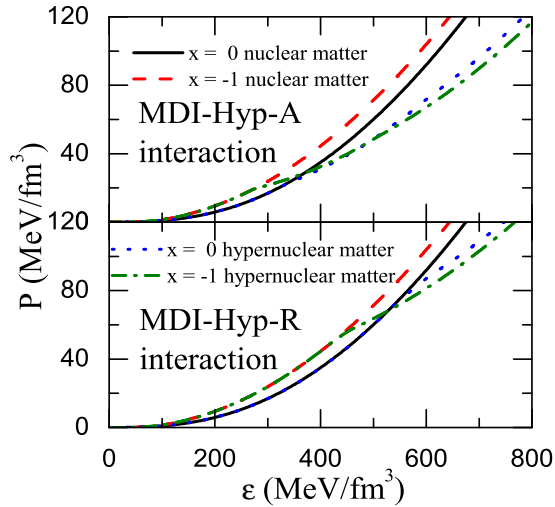}}
\caption{(Color online) Equation of state of pure
nuclear matter and hypernuclear matter from the MDI-Hyp interaction
with $x=0$ and $x=-1$ as well as with attractive (upper panel) and
repulsive (lower panel) $\Sigma$N interactions.}\label{EOSH}
\end{figure}

In Fig.~\ref{EOSH}, we show the equation of state of nuclear matter
from the MDI interaction with $x=0$ and $x=-1$ and of hypernuclear
matter from the MDI-Hyp-A and MDI-Hyp-R interactions. It is seen
that the EOS is softened when hyperons are present as compared to
the EOSs of pure nuclear matter. We note that the EOS at moderate
densities plays an important role in determining the maximum mass of
neutron stars. Without hyperons, the EOS at moderate densities is
stiffer for smaller value of $x$, while it becomes stiffer for
larger value of $x$ when hyperons are included, especially for a
repulsive $\Sigma$N interaction. As shown in Fig.~\ref{MDIHPF}, a
stiffer symmetry energy leads to a larger number of hyperons, which
results in a softer EOS as a result of the lower pressure due to the
presence of more degrees of freedom, and the effect of hyperons on
the EOS is smaller for a soft symmetry energy, as fewer hyperons are
then present in the hypernuclear matter. One the other hand, the
symmetry energy contribution to the pressure of the hypernuclear
matter is larger for $x=-1$ than $x=0$. As a result, the EOS of
hypernuclear matter shows similar stiffness for different symmetry
energy parameters. Furthermore, the EOS from the MDI-Hyp-R
interaction is stiffer than that from the MDI-Hyp-A interaction, as
$\Sigma$ particles do not appear in the former case as shown in
Fig.~\ref{MDIHPF}. Our results thus show that the hypernuclear
matter has a stiffer (softer) EOS at lower densities but a softer
(stiffer) EOS at moderate densities with a stiffer (softer) nuclear
symmetry energy. We note that the results from both $x=0$ and $x=-1$
are consistent with the constraints on the nuclear equation of state
obtained from the analysis of the collective flow data in heavy-ion
collisions~\cite{Dan02}.

\subsection{The hadron-quark phase transition}

\begin{figure}[h]
\centerline{\includegraphics[scale=0.9]{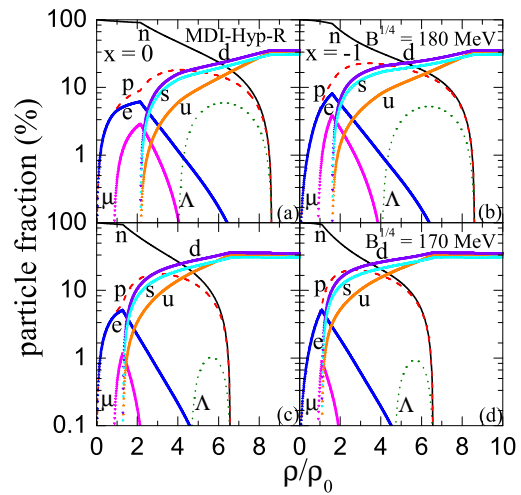}} \caption{(Color
online) Particle fractions in a hypernuclear matter with the
presence of hadron-quark phase transition from the MDI-Hyp-R
interaction with $x=0$ ((a) and (c)) and $x=-1$ ((b) and (d)) for
the hadron phase and the MIT bag model for the quark phase. Results
from $B^{1/4}=180$ MeV ((a) and (b)) and $170$ MeV ((c) and (d)) are
shown for comparison. }\label{MDIHQPTPF}
\end{figure}

At higher densities in the core of a neutron star, a transition from
hadron matter to quark matter is expected to occur. Here we only
consider the hadron-quark phase transition for the repulsive
$\Sigma$N interaction as it is more consistent with the latest
empirical
information~\cite{Batty94,Mares95,Noumi02,Har05,Friedman07}. In
Fig.~\ref{MDIHQPTPF}, we display the particle fractions of each
species in the presence of a hadron-quark phase transition, with the
hadron phase from the MDI-Hyp-R interaction with $x=0$ and $x=-1$
and the quark phase from the MIT bag model with bag constants
$B^{1/4}=180$ MeV and $170$ MeV. The particle fractions are weighted
by the baryon number and corresponding phase fraction. It is seen
that for $B^{1/4}=180$ MeV, the phase transition begins at the
density $0.35$ fm$^{-3}$ and ends at the density $1.38$ fm$^{-3}$
for $x=0$, and begins at the density $0.26$ fm$^{-3}$ and ends at
the density $1.38$ fm$^{-3}$ for $x=-1$, while for $B^{1/4}=170$ MeV
it begins at the density $0.21$ fm$^{-3}$ and ends at the density
$1.05$ fm$^{-3}$ for $x=0$, and begins at the density $0.18$
fm$^{-3}$ and ends at the density $1.05$ fm$^{-3}$ for $x=-1$. The
hadron-quark phase transition thus happens at lower baryon number
density for a stiffer symmetry energy and for a smaller value of
$B$, while the density at which the hadron-quark phase transition
ends seems to depend only on the value of $B$ but not much on the
value of the symmetry energy parameter $x$. With a smaller value of
the bag constant, the hadron-quark phase transition both begins and
ends earlier. It is also seen that $d$ and $s$ quarks occupy a
larger fraction than $u$ quarks in the mixed phase because of their
negative charges, so the lepton fraction decreases while the proton
fraction increases when the hadron-quark phase transition occurs.
Furthermore, only $\Lambda$ hyperons (no other hyperons) appear in
the mixed phase in our model. The fraction of $\Lambda$ hyperons is,
however, sensitive to the value of the bag constant $B$, and with a
smaller value of $B$ its fraction becomes smaller.

\begin{figure}[h]
\centerline{\includegraphics[scale=0.9]{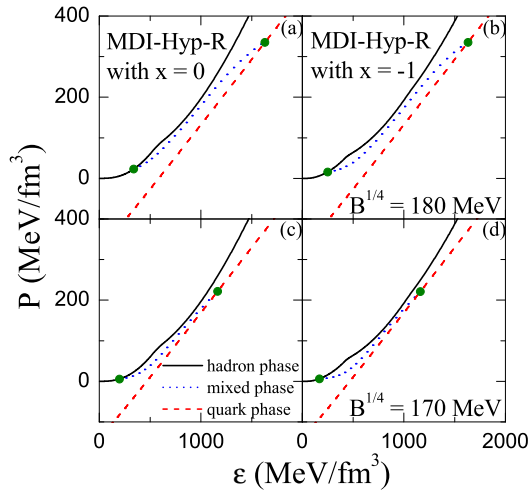}} \caption{(Color
online) Same as Fig.~\ref{MDIHQPTPF} for the equation of
state.}\label{EOSHQPT}
\end{figure}

For the EOSs in the presence of the hadron-quark phase transition,
they are shown in Fig.~\ref{EOSHQPT} with $B^{1/4}=180$ MeV and 170
MeV. Our results show that the EOSs of the mixed phase and the quark
phase are both softened in comparison with a pure hadron phase. The
difference between the results from $x=0$ and $x=-1$ is not large
except for the different starting density of the hadron-quark phase
transition. Since the EOS is more sensitive to the value of bag
constant than to the value of $x$ when the quark degrees of freedom
are introduced, the bag constant $B$ is thus the main parameter in
determining the EOS of dense matter as reported in other
work~\cite{Pan07}. The energy density at the end of the hadron-quark
phase transition is about $1.2$ GeV/fm$^3$ for $B^{1/4}=180$ MeV and
about $1.6$ GeV/fm$^3$ for $B^{1/4}=170$ MeV, with the former closer
to the value of about $1$ GeV/fm$^3$ obtained from the lattice QCD
calculation~\cite{Cheng10} and extracted from heavy-ion collision
experiments~\cite{Heinz00}. Compared with the equation of state
constrained by the collective flow data in heavy-ion
collisions~\cite{Dan02}, our results from both values of bag
constants satisfy the empirical constraint.

\subsection{Hybrid stars}

In this subsection, we use the MDI-Hyp interaction to study
the properties of static hybrid stars with spherically symmetric
mass distributions. In particular, we calculate the mass-radius
relation of a hybrid stars using Tolman-Oppenheimer-Volkoff (TOV)
equation~\cite{Oppen,Morrison}
\begin{equation}\label{tov}
\frac{d P}{d r} = -\frac{(\epsilon+P)(m_g+4 \pi r^3 P)}{r(r-2m_g)},
\end{equation}
where $m_g$ is the gravitational mass inside the
radius $r$ of the hybrid star given by
\begin{equation}\label{m}
\frac{dm_g}{dr} = 4\pi r^2\epsilon(r),
\end{equation}
with $\epsilon(r)$ being the energy density. Above equations are solved by
starting from a central energy density $\epsilon(r=0)\equiv
\epsilon_c$ and integrating outward until the pressure on the
surface of the hybrid star defined by $r=R$ vanishes, i.e, $P(R)=0$.
This gives the radius $R$, and the total gravitational mass of a
hybrid star is then given by $M=m_g(R)$.

In our calculations, the hybrid star is divided into three parts
from the center to the surface: the liquid core, the inner crust and
the outer crust as in our previous work~\cite{XCLM09a,XCLM09b}. In
the liquid core it is assumed to be the hypernuclear
matter or that with the hadron-quark phase transition, and the
resulting EOSs shown in previous subsections are used. In the inner
crust, a parameterized EOS of $P=a+b \epsilon^{4/3}$ is used as in
the previous treatment~\cite{XCLM09a,XCLM09b}. The outer crust
usually consists of heavy nuclei and the electron
gas, where we use the BPS EOS~\cite{BPS}. The transition density
$\rho_{\rm t}$ between the liquid core and the inner crust has been
consistently determined in our previous work~\cite{XCLM09a,XCLM09b},
and for the density which distinguishes the edge of inner crust and
outer crust, we take it to be $\rho_{\rm out}=2.46\times10^{-4}$
fm$^{-3}$. The parameters $a$ and $b$ can then be determined by the
pressures and energy densities at $\rho_{\rm t}$ and $\rho_{\rm
out}$.

\begin{figure}[h]
\centerline{\includegraphics[scale=0.9]{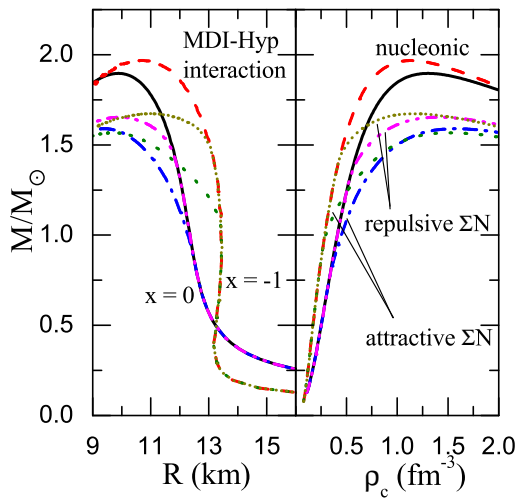}} \caption{(Color
online) The hybrid star mass as a function of radius (left panel)
and central density (right panel) based on the MDI-Hyp interaction
with $x=0$ and $x=-1$. Results from attractive and repulsive
$\Sigma$N interactions are shown for comparison. Results from a pure
nucleonic approach are also displayed.}\label{MRrhocH}
\end{figure}

In Fig.~\ref{MRrhocH}, we show the mass-radius (M-R) and
mass-central density (M-$\rho_{\rm c}$) relations of hybrid stars
from the EOS of the hypernuclear matter without the hadron-quark
phase transition as obtained above based on the MDI-Hyp interaction.
Results using both attractive and repulsive $\Sigma$N interactions
are shown, and the results from a pure nucleonic approach without
hyperons are also shown for comparison. The maximum mass obtained
with hyperons in hybrid stars is seen to be smaller as the EOS of
the hypernuclear matter is softer than that of the nuclear matter.
The EOSs at moderate densities play an important role in determining
the maximum mass, as the M-$\rho_{\rm c}$ relation shows that the
neutron star mass can not increase even the central density reaches
a very high value. The mass is larger when a repulsive $\Sigma$N
interaction is used as a result of the smaller number of degrees of
freedom and stiffer EOS, as discussed in the previous subsection.
For smaller values of $\rho_c$ a stiffer symmetry energy gives a
larger mass as the nuclear matter dominates this density range, and
the results are thus the same as those in our previous
work~\cite{XCLM09a,XCLM09b}. With larger $\rho_c$, the effect due to
the nuclear symmetry energy is small, and the mass from $x=0$ is
similar to that from $x=-1$ because of the opposite effects from the
symmetry pressure and the fraction of hyperons on the equation of
state. The maximum mass obtained with an attractive $\Sigma$N
interaction is $1.59 M_{\odot}$ for $x=0$ and $1.57 M_{\odot}$ for
$x=-1$, where $M_{\odot}$ is the solar mass. With a repulsive
$\Sigma$N interaction the maximum mass increases to $1.65 M_{\odot}$
for $x=0$ and $1.67 M_{\odot}$ for $x=-1$. Our results thus differ
from those of many previous works~\cite{Baldo00,Vidana00} that the
maximum mass of hybrid stars cannot reach the canonical value of
$1.4 M_{\odot}$. In our model, the radius of a hybrid star with a
mass of $1.4 M_{\odot}$ is $11.2$ km for $x=0$ and $11.9$ km for
$x=-1$ for an attractive $\Sigma$N interaction, and $11.9$ km for
$x=0$ and $13.2$ km for $x=-1$ for a repulsive $\Sigma$N
interaction, respectively.

\begin{figure}[h]
\centerline{\includegraphics[scale=0.9]{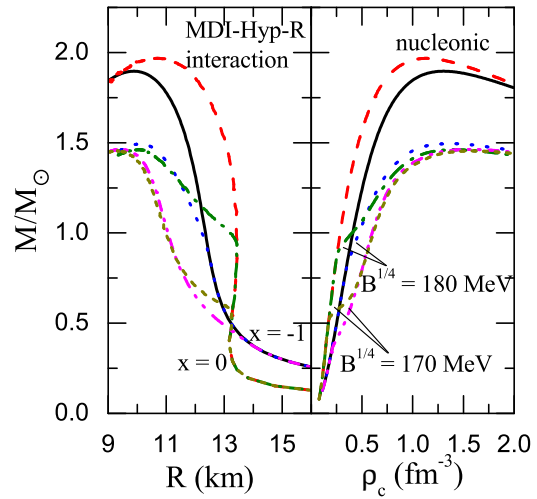}} \caption{(Color
online) The hybrid star mass as a function of radius (left panel)
and central density (right panel) in the presence of the
hadron-quark phase transition. Results from the MDI-Hyp-R
interaction for the hadron phase with $x=0$ and $x=-1$ and the MIT
bag model for the quark phase with $B^{1/4}=180$ MeV and $170$ MeV
are shown for comparison. Results from a pure nucleonic approach are
also displayed.}\label{MRrhocHQPT}
\end{figure}

In Fig.~\ref{MRrhocHQPT}, the M-R and M-$\rho_c$ relations are
displayed for hybrid stars with the hadron-quark phase transition in
their liquid core. Again, results from the hadron phase with $x=0$
and $x=-1$ and the quark phase with $B^{1/4}=180$ MeV and
$B^{1/4}=170$ MeV are shown, and those from a pure nucleonic
approach are also shown for comparison. The maximum mass for
$B^{1/4}=180$ MeV is $1.50 M_{\odot}$ for $x=0$ and $1.46 M_{\odot}$
for $x=-1$, and for $B^{1/4}=170$ MeV it is $1.46 M_{\odot}$ for
$x=0$ and $1.45 M_{\odot}$ for $x=-1$, respectively. The radius of a
standard neutron star with a mass of $1.4 M_{\odot}$ is $11.0$ km
for $x=0$ and $10.8$ km for $x=-1$ for $B^{1/4}=180$ MeV, and $10.2$
km for $x=0$ and $10.0$ km for $x=-1$ for $B^{1/4}=170$ MeV. If we
further reduce the value of $B$, the hadron-quark phase transition
would happen at an even lower density, and the maximum mass for the
hybrid star would correspond to a larger central density and a
smaller radius. Above results are obtained with the Gibbs
construction for the hadron-quark phase transition. With the Maxwell
construction, the radius of a maximum mass hybrid star would be
larger. Within our model, both treatments of the phase transition
give, however, reasonable masses and radii for hybrid stars.

Finally, we note that the original MDI interaction does not lead to
the violation of causality in $\beta$-stable $npe\mu$ matter up to
$10\rho_0$ as shown in Panel (d) of Fig.~2 in Ref.~\cite{XCLM09b}.
Since the EOS of neutron star matter is softened by the presence of
hyperons and quarks, the causality condition is still satisfied for
the extended MDI interaction used in the present study, and this is
confirmed by explicit calculations.

\section{Summary}\label{summary}

We have extended the MDI interaction for the nucleon-nucleon
effective interaction in nuclear medium to include the
nucleon-hyperon and hyperon-hyperon interactions by assuming that
they have the same density and momentum dependence as that for the
nucleon-nucleon interaction. The parameters were determined by
fitting the empirical hyperon single-particle potentials in
symmetric nuclear matter at its saturation density. As an example
for the application of the extended MDI interaction, we investigated
the properties of hybrid stars that include not only the hyperon
degree of freedom but also that of quarks by taking into account the
phase transition between the hadron and quark phases. Our results
indicate that the extended MDI interaction can give a reasonable
description of the hypernuclear matter. We found that the EOS of the
hypernuclear matter is much softer than that of the pure nuclear
matter, and that it becomes even softer if the hadron-quark phase
transition is included. The masses and radii of hybrid stars were
also studied with these EOSs, and they were found to remain
reasonable after including hyperons and the hadron-quark phase
transition. We have also studied the effects of attractive and
repulsive $\Sigma$N interactions and different symmetry energies on
the hybrid star properties. The results show that the appearance of
the $\Sigma$ hyperon in hybrid stars depends sensitively on the sign
of the $\Sigma$N interaction with a repulsive $\Sigma$N interaction
giving a higher critical density for the appearance of $\Sigma$
hyperons. In addition, a stiffer symmetry energy usually leads to a
larger fraction of hyperons in the hypernuclear matter. We further
found that both the low-density boundary of the hadron-quark phase
transition and the EOS at high densities in hybrid stars are more
sensitive to the bag constant than to the stiffness of the nuclear
symmetry energy at high densities. This extended MDI interaction,
which gives an isospin- and momentum-dependent in-medium effective
interactions for the baryon octet, will also be very useful in
transport models that simulate heavy-ion reactions in future
radioactive beam facilities, particularly at the FAIR/GSI energies.

\begin{acknowledgments}

We thank David Blaschke for critical comments on an earlier version of
this paper. This work was supported in part by U.S. National Science Foundation
under Grant No. PHY-0758115, PHY-0652548 and PHY-0757839, the Welch
Foundation under Grant No. A-1358, the Research Corporation under
Award No. 7123, the Texas Coordinating Board of Higher Education
Award No. 003565-0004-2007, the National Natural Science Foundation
of China under Grant Nos. 10675082 and 10975097, MOE of China under
project NCET-05-0392, Shanghai Rising-Star Program under Grant No.
06QA14024, the SRF for ROCS, SEM of China, the National Basic
Research Program of China (973 Program) under Contract No.
2007CB815004.

\end{acknowledgments}


\begin{thebibliography}{99}
\bibitem{footnote}This convention is different from that of
Ref.~\cite{LCK08} which uses $\tau_N=1/2$ for neutron and
$\tau_N=-1/2$ for proton.
\end{thebibliography}
\end{document}